\documentclass[12pt,preprint]{aastex}    
\input{psfig.sty}
\input
\usepackage{color}

\newcommand{\be}{\begin{equation}}
\newcommand{\ee}{\end{equation}}

\def\no{\noindent}

\newcommand{\ifm}[1]{\relax\ifmmode#1\else$\mathsurround=0pt #1$\fi}
\newcommand{\kms}{\ifmmode\,{\rm km}\,{\rm s}^{-1}\else km$\,$s$^{-1}$\fi}

\newcommand{\ltsima}{$\; \buildrel < \over \sim \;$}
\newcommand{\lsim}{\lower.5ex\hbox{\ltsima}}
\newcommand{\gtsima}{$\; \buildrel > \over \sim \;$}
\newcommand{\gsim}{\lower.5ex\hbox{\gtsima}}

\def\M11{M_{11}}
\def\V100{V_{100}}
\def\R1{R_{Mpc}}
\def\T6{T_6}

\begin{document}

\title{Galaxy luminosity function and Tully-Fisher relation: reconciled through rotation-curve studies}
\shorttitle{Galaxy luminosity function and Tully-Fisher relation}

\author { Andrea Cattaneo\altaffilmark{1}, Paolo Salucci\altaffilmark{2},  \& Emmanouil Papastergis\altaffilmark{3}}

\altaffiltext{1}{Laboratoire d'Astrophysique de Marseille, UMR 6110 CNRS, Univ. d'Aix-Marseille, 38 rue F. Joliot-Curie, 13388 Marseille cedex 13, France,  {\textit{r-mail:}} andrea.cattaneo@oamp.fr}

\altaffiltext{2}{Department of Astrophysics, SISSA, Via Beirut, 2-4, 34014 Trieste, Italy, {\textit{e-mail:}} salucci@sissa.it}

\altaffiltext{3}{Center for Radiophysics and Space Research, Space Sciences Building,
Cornell University, Ithaca, NY 14853, USA, {\textit{e-mail:}} papastergis@astro.cornell.edu}

\begin{abstract}
The relation between galaxy luminosity $L$ and halo virial velocity $v_{\rm vir}$ required to fit the galaxy luminosity function differs from the observed Tully-Fisher relation between
$L$ and disc  speed $v_{\rm rot}$. Hence the problem of reproducing the galaxy luminosity function and the Tully-Fisher relation simultaneously has plagued semianalytic models since their inception. Here we study the relation between $v_{\rm rot}$ and $v_{\rm vir}$ by fitting observational average rotation curves of disc galaxies binned in luminosity.
We show that the $v_{\rm rot}$- $v_{\rm vir}$ relation that we obtain in this way can fully account for this seeming inconsistency. Therefore, the reconciliation of the 
luminosity function with the Tully-Fisher relation rests on the complex dependence of $v_{\rm rot}$ on $v_{\rm vir}$, which arises because
the ratio of stellar mass to
dark matter mass is a strong function of halo mass.

\end{abstract}

\maketitle

\label{firstpage}



\section{Introduction}

In the standard $\Lambda$CDM cosmology, $\sim 84\%$ of the mass of the Universe is composed of dark matter (DM), which dominates gravitational evolution on large scales. Galaxies are baryonic condensations at the centres of DM haloes formed by gravitational instability of primordial density fluctuations. The main goal for studies of galaxy formation in a cosmological context
is to explain the properties of galaxies (luminosities, spectra, sizes, and morphologies) in terms of the growth histories of their host haloes. 

Historically, the main observational constraints on the link between luminous galaxies and the underlying DM distribution come from the galaxy luminosity function (LF) and from the Tully-Fisher (TF) relation $L\propto v_{\rm rot}^\eta$  that links the luminosity $L$ of a disc galaxy and its reference  speed $v_{\rm rot}$ ($\eta\sim3-4$ depending on the band. Fig.~1 shows the 
$I$-band TF relation of \citealp{yegorova_salucci07}).
Today, there are further constraints from weak lensing data \citep{mandelbaum_etal06, leauthaud_etal10,reyes_etal12},
velocity dispersion profiles \citep{martinsson_etal13}
\footnote{One can measure  the absorption lines of disc  stars at some height $z$ over the disc and derive a stellar velocity dispersion profile,
which can be used for global mass modelling.
This method is in principle very useful,  especially for spirals in which the rotation curves (RCs) alone  are not sufficient for it.
In practice, recent work \citep{martinsson_etal13}
shows that this technique is not yet able to give a unique luminous-DM decomposition but that, coupled  with other determinations, it can be decisive to improve their accuracy.}, 
and kinematics of satellite galaxies \citep{conroy_etal07,more_etal11}. The latter probe the gravitational potential of spiral galaxies at large radii.
\citet{yegorova_etal11} find that their results are consistent with those from rotation-curve (RC) data in a statistical sense.

The importance of the LF and the TF relation for constraining the relation between galaxy luminosity and halo mass is readily explained. 
Let us start from the LF.
If a volume of the universe contains $n$ haloes of mass $M_h$, and if each halo of mass $M_h$ hosts a galaxy of luminosity $L$, then the volume must contain $n$ galaxies with luminosity $L$. More generally, the number of galaxies brighter than $L$ in a given band must be equal to the number of haloes more massive than $M_h$ if $L$ is a growing function of $M_h$ - a general predictions of galaxy formation models and simulations in bands where luminosity traces stellar mass. The LF of galaxies is observed and the $\Lambda$CDM model makes strong predictions for the mass function of DM haloes. Hence, there is a well defined 
$L - M_h$ relation that cosmological models must satisfy to fit the galaxy LF.
It is straightforward to convert
this relation into a relation between $L$ and $v_{\rm vir}$ by using the fitting formulae of \citet{bryan_norman98} for the virial overdensity contrast.

The  black dashed curve in Fig.~1 shows the $L-v_{\rm vir}$ relation that we obtain when we apply 
this method to
the  LF of \citet{papastergis_etal12} and to the halo mass function for a $\Lambda$CDM cosmology
with $h_0 = 0.73$, $\Omega_m=0.24$, $\Omega_\Lambda=0.76$, $\sigma_8=0.76$.
The $I$-band LF of \citet{papastergis_etal12} is constructed by computing the Kron-Cousin $I$-band magnitude of each individual galaxy from its SDSS $r$- and $i$-band magnitudes.
We have verified that this LF is consistent with the SDSS DR6 i-band LF  of \citet{monterodorta_prada09} when we convert the latter to $I$ band by using mean $I-i$ colours in
\citet{fukugita_etal95}.
The halo mass function comes a cosmological N-body simulation that was run by the Horizon Project ({\tt http://www.projet-horizon.fr})
and is the same that we used in \citet{papastergis_etal12}, to whom we refer for all details concerning the abundance matching procedure.

\begin{figure}
\begin{center}
\includegraphics[scale=0.75]{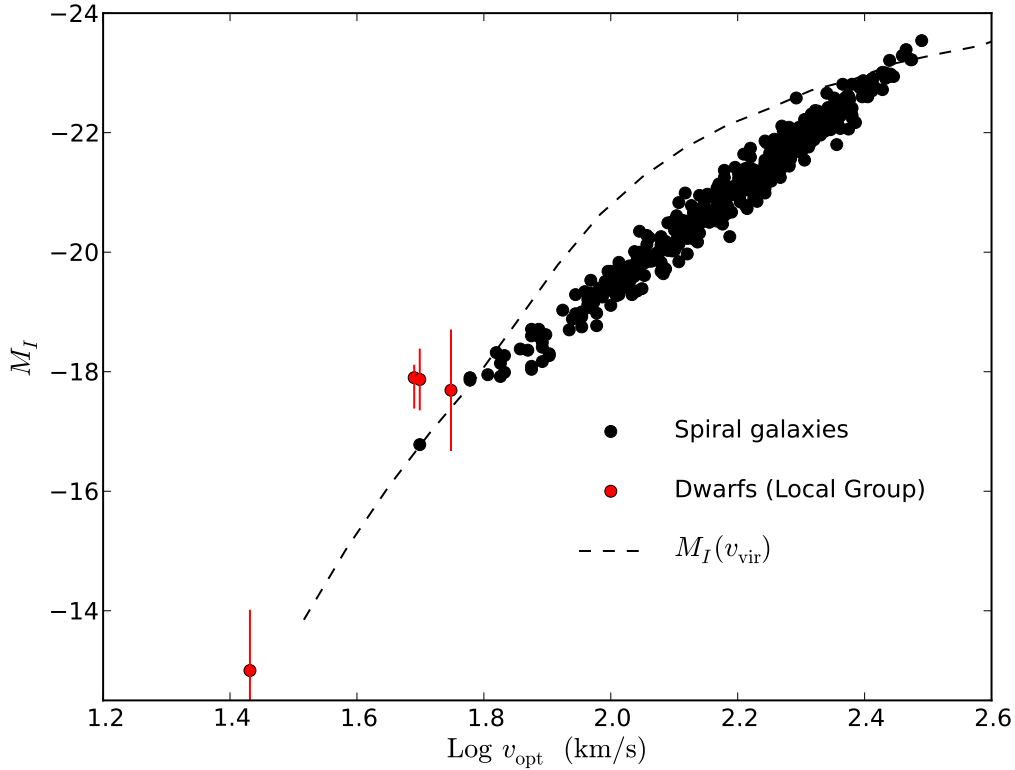} 
\end{center}
\label{stripping_vs_radius}
\caption{
 Black symbols: the TF relation for the spiral-galaxy sample of \citet{yegorova_salucci07}. $M_I$ is the absolute magnitude in the $I$-band
of Kron-Cousin, $v_{\rm opt}$ is the  speed at the optical radius (equal to 3.2 exponential radii).
Red symbols: four dwarf galaxies from \citet{salucci_etal12} inserted to extend the TF relation to $M_I\sim -13$.
Black dashed curve: the $L_I$ - $v_{\rm vir}$ relation that we obtain by matching the Kron-Cousin $I$-band LF of \citet{papastergis_etal12} and the halo mass function from a cosmological simulation with $h_0 = 0.73$, $\Omega_m=0.24$, $\Omega_\Lambda=0.76$, 
and $\sigma_8=0.76$.}
\end{figure}

Let us now consider the TF relation. The disc rotation speed $v_{\rm rot}(r)$ measures the total mass $M$ within radius $r$,  which is dominated by DM for large values of $r$. Since, at large radii, the RCs of disc galaxies are nearly flat, the TF relation implies a relation between $L$ and $v_{\rm vir}$.
The trouble is that this relation differs from the one that we find from the LF.

Fig.~1 illustrates the problem by comparing the  $L_I-v_{\rm vir}$  relation that we derive from abundance matching (black dashed line) with the $I$-band TF relation of Yegorova \& Salucci (2007, black symbols).  This discrepancy cannot be attributed to variations in mass-to-light ratio because the entire analysis has been done using the Kron-Cousin 
$I$-band luminosity both for the LF and the TF relation. This is a strong point of our work and the reason why Fig.~1 is totally independent of any assumption on the stellar
mass-to-light ratio (on the issue of mass-to-light ratios, also see \citealp{portinari_salucci10}).

The discrepancy shown in Fig.~1 has been known for twenty years and is independent of the abundance-matching method because any model that matches the galaxy LF ends up with an $L_I-v_{\rm vir}$ relation  similar to the black dashed line. 
Either models are calibrated on the TF relation and fail to fit the LF \citep{kauffmann_etal93} or they are calibrated on the LF and they fail to reproduce the TF relation \citep{heyl_etal95}.
This inconsistency has plagued galaxy formation models since the earliest studies. The discrepancy persists today \citep{guo_etal10}.

\citet{dutton_etal10} and \citet{reyes_etal12} have recently pointed that the problem derives from the incorrect assumption that $v_{\rm rot}$ is a good tracer of $v_{\rm vir}$. \citet{dutton_etal10} used different methods (satellite kinematics, weak lensing observations, abundance matching) to investigate the halo masses of early- and late-type galaxies and compared the stellar mass - virial velocity relations derived from these studies with the Faber-Jackson and the TF relation. They found that 
the disc rotation speed $v_{\rm rot}$ and the virial velocity $v_{\rm vir}$ are systematically different and that $v_{\rm rot}/v_{\rm vir}$ varies with stellar mass.
\citet{reyes_etal12} did the same type analysis for a galaxy sample for which they had both the TF relation and weak lensing data.
Their results are in qualitative agreement with \citet{dutton_etal10}, though their values of $v_{\rm rot}/v_{\rm vir}$ are systematically higher.

Our article follows the same general philosophy of these two previous studies but the method is completely different because first we determine $v_{\rm rot}/v_{\rm vir}$
as a function of $L_I$ from the modelling of disc RCs and then we use this result to convert the $L_I-v_{\rm vir}$  relation from abundance matching into
a relation between $L_I$ and $v_{\rm rot}$, which can be compared with the TF relation.
The originality of the article is that, following \citet{salucci_etal07}, we compute the $v_{\rm rot}/v_{\rm vir}$-$L_I$ with a method that is completely independent of the galaxy LF {\it and} we find that this relation is precisely the one we need to reconcile the TF relation with the galaxy LF.

The structure of the article is as follows. In Section~2, we present our RC analysis and our results for the $v_{\rm rot}$-$v_{\rm vir}$ relation.
In Section~3, we combine the results of the previous section with those from abundance matching to compute a TF that will be found to be in agreement with the observations, and
we discuss the significance of this final result.

\section{The  relation between $v_{\rm opt}$ and $v_{\rm vir}$ }

Our method to compute $v_{\rm vir}$ is conceptually quite simple.
We fit the RCs of spiral galaxies by assuming that they are made of two components: a baryonic disc and a DM halo.
The best-fit DM-halo profile is extrapolated out the radius $r_{\rm vir}$ where the mean density equals the critical density of the Universe times the virial overdensity
contrast computed with the formulae by  \citet{bryan_norman98}. The virial velocity $v_{\rm vir}$ is equal to the circular velocity at this radius.
In practice, to apply this method, we need to specify three things: a model for the density distribution of the baryonic disc, a model for the density distribution of the DM haloes, and the RCs on which we intend to do the analysis. Let us analyse these three elements one by one.

\subsection{The baryonic disc model}
We model the baryonic disc with a single exponential profile. This model contains two parameters: the total disc mass $M_d$ (stars plus gas) and the disc exponential radius $r_d$. We do not treat $r_d$ as a free parameter of the fit. Instead, we require it to be equal to the exponential radius of the
$I$-band surface-brightness profile. This is equivalent to assuming that the gas and the stellar disc have the same scale-length.
In fact, the gas distribution is usually more extended, but this has almost no effect on our results for the following reason. Small galaxies are completely DM-dominated.
An error on the spatial extension of the baryonic component will have little effect on the best-fit parameters for the dominant DM component 
(this point is shown quantitatively in \citealp{persic_etal96}).
In massive galaxies, the baryonic component is dynamically important, but these galaxies have low gas fractions. Therefore, the gas component has a negligible effect on the size of the baryonic disc.

\subsection{The dark matter halo model}

Cosmological simulations of dissipationless hierarchical clustering in a cold DM Universe find that the density distribution of DM haloes are described by the
NFW \citep{navarro_etal97} profile:
\begin{equation}
\rho(r) = {\rho_0\over {r\over r_0}\left(1+{r\over r_0}\right)^2},
\end{equation}
with concentration
\begin{equation}
c(M_{\rm vir})={r_{\rm vir}\over r_0}=9.6\left({M_{\rm vir}\over 10^{12}h^{-1}M_\odot}\right)^{-0.075}
\end{equation}
for haloes at $z\simeq 0$ \citep{klypin_etal11}.
This profile provides a poor fit to the RCs of spiral galaxies in the inner regions (\citealp{flores_primack94}; and \citealp{moore94}; but also see
\citealp{swaters_etal03}). 

The \citet{burkert95} profile
\begin{equation}
\rho(r) = {\rho_0\over \left(1+{r\over r_0}\right)\left(1+{r^2\over r_0^2}\right)},
\end{equation}
gives a much better fit to the observed RCs (Fig.~2), though with concentration-parameter values that are $\sim 1.4-1.5$ times larger than suggested by Eq.~(2).
As this is is a phenomenological paper, the fact that the Burkert model fits the observation is a good enough reason for using it independently of any theoretical argument.
However, some reader may wonder
whether it is self-consistent to use the Burkert profile (Eq.~3) alongside the halo mass function from a simulation that assumes a cold DM cosmology.
To address this objection, we remark that 
our choice is justified for two reasons. First, Eq.~(1) and Eq.~(3) are almost identical at $r\gsim r_0$ and differ only at small radii, while here we are interested in the DM density distribution at large radii. Recent hydrodynamical simulations have shown that the difference at small radii may be understood as an
effect of supernova feedback, which was not considered in pure N-body simulations \citep{governato_etal10,pontzen_governato12,brook_etal12,teyssier_etal13}.
Second, the concentration difference can be explained as being due to adiabatic contraction. We have verified this point quantitatively by comparing the DM density profile in a simulation by \citet{geen_etal13} in the cases with pure DM and with baryon cooling.

\subsection{The RCs}
Our analysis is based on a sample of 967 spiral galaxies for which there are high-quality H$\alpha$ data and $I$-band photometry 
\citep{persic_salucci95,mathewson_etal92}. The sample is split in eleven luminosity bin and each luminosity bin is analysed separately.
Instead of analysing the RC of each galaxy individually and then computing an average $v_{\rm rot}/v_{\rm vir}$, we directly analyse the co-added
RCs of PSS \citep{persic_etal96}, obtained by stacking the RCs of the galaxies in each bin. 
Following PSS, the co-added RCs are computed by averaging the values of
$v_{\rm rot}$ in bins of $r/r_{\rm opt}$, where $r_{\rm opt}=3.2r_d$ ($r_d$ is the exponential radius of the $I$-band surface brightness profile).
The points with error bars in Fig.~2 show the co-added RCs in the eleven $I$-band magnitude bins.

\begin{figure*}
\begin{center}
\includegraphics[scale=0.75]{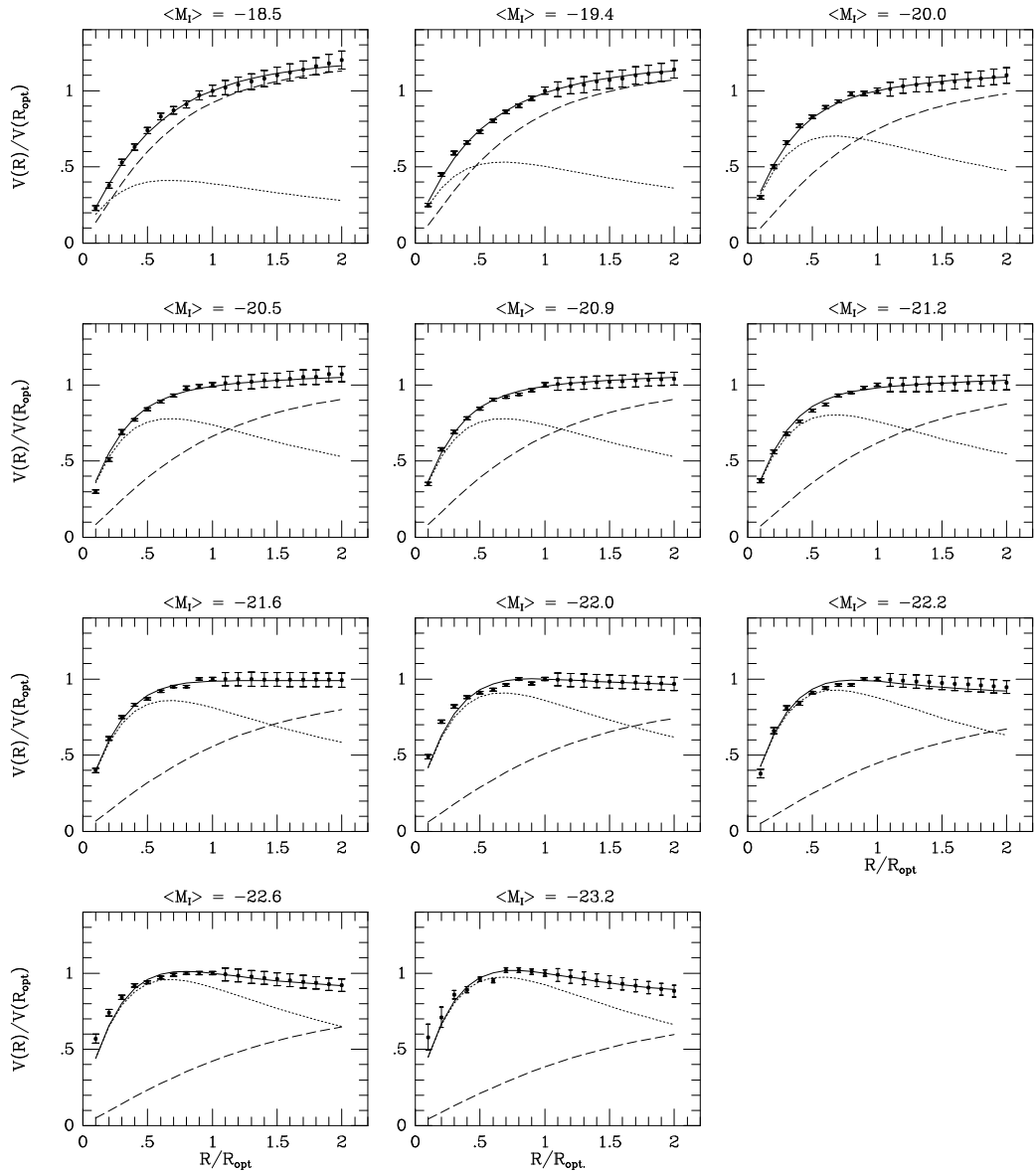} 
\end{center}
\caption{Points with error bars: the co-added RCs in eleven $I$-band magnitude bins for the 967 spiral galaxies in PSS's sample.
Curves: each co-added RC is fitted with a disc (dotted curve) and a halo (dashed curve) contribution. Their sum is shown by the solid curve.
Faint spirals are dark-matter dominated at all radii.
Bright spirals are disc-dominated at all radii.}
\end{figure*}

\subsection{The fit}

In each magnitude bin, we fit the co-added RC with a disc plus halo model (Fig.~2).
We do the fit at $0.5<r/r_{\rm opt}<2$ to give more weight to the outer regions.
The result is generally quite good (compare the solid lines and the points with error bars).

The disc contribution is a strong function of luminosity.
Faint galaxies ($M_I\gsim 19$) are entirely DM-dominated. Their RCs rise steeply out to $r=2r_{\rm opt}$.
Bright spirals ($M_I<-22$) are disc dominated out to very large radii. Their RCs peak at $0.5<r/r_{\rm opt}<1$ and decrease at larger radii.

Let $v_{\rm opt}$ be the mean rotation speed at the optical radius and let $v_{\rm vir}$ be the virial velocity obtained for the best-fit halo parameter
by extrapolating the halo contribution out the virial radius (the radius within which the mean density equals the critical density of the Universe times the critical overdensity contrast computed with
the fitting formulae of \citealp{bryan_norman98}).
By computing $v_{\rm opt}$ and $v_{\rm vir}$ for each of our magnitude bins, we obtain the 
$v_{\rm opt}$ - $v_{\rm vir}$ relation that is shown by the red solid curve in Fig.~3.

\begin{figure*}
\begin{center}
\includegraphics[scale=0.75]{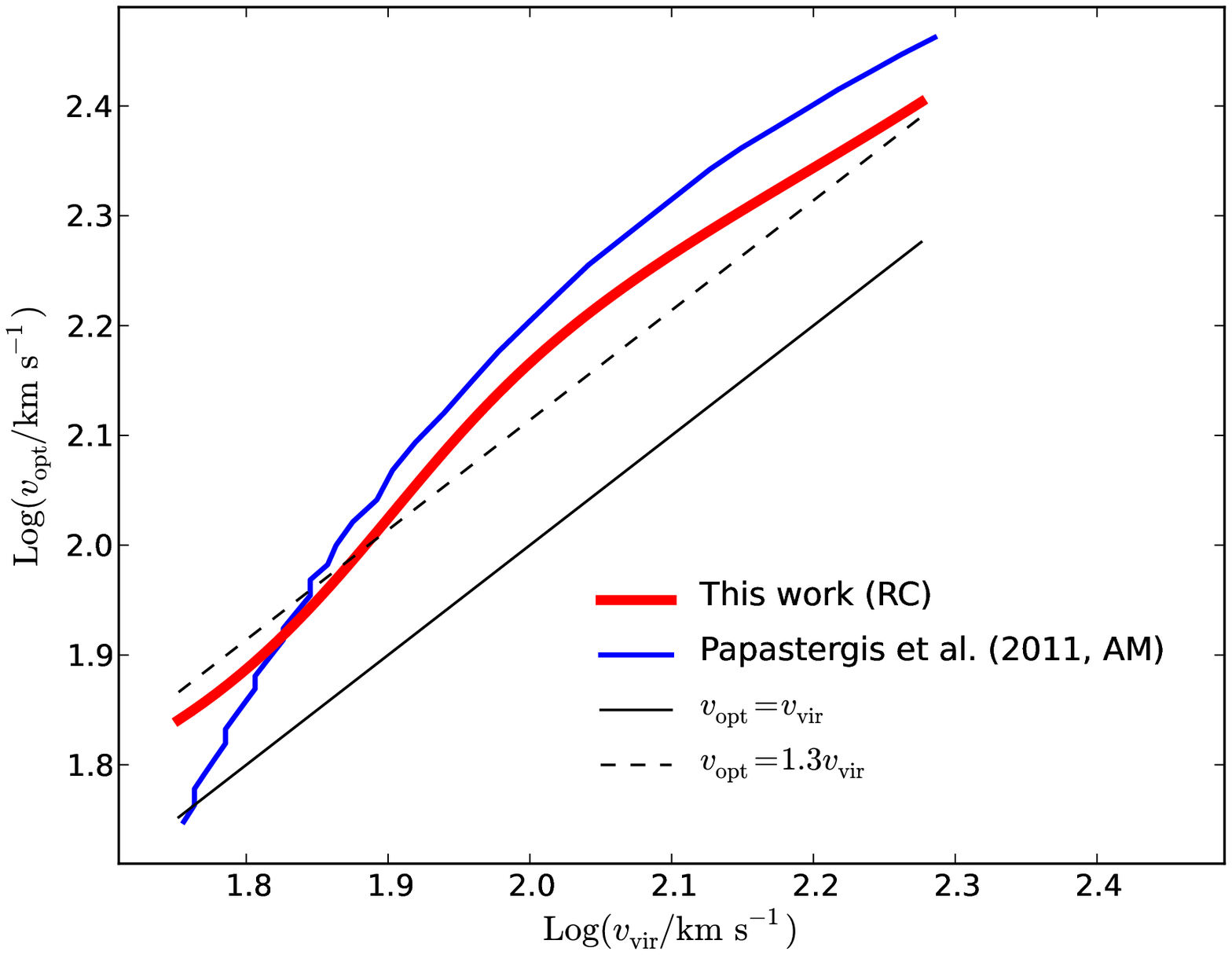} 
\end{center}
\caption{Thick red curve: the $v_{\rm opt}$ - $v_{\rm vir}$ relation that we obtain from our rotation curve (RC) analysis.
Blue curve: the $v_{\rm opt}$ - $v_{\rm vir}$ that \citet{papastergis_etal11} obtained with the abundance matching (AM) method by matching the 
galaxy velocity function and the halo virial velocity function.
Thin black solid line: the relation that one would have if $v_{\rm opt}=v_{\rm vir}$ (shown for comparison).
Thin black dashed line: the relation that one would have if $v_{\rm opt}=1.3v_{\rm vir}$ (shown for comparison).}
\end{figure*}

Fig.~3 shows that $v_{\rm opt}$ differs from $v_{\rm vir}$. The difference is by a factor of $\sim 1.3$ but its precise value varies with $v_{\rm vir}$ and has a 
maximum of $v_{\rm opt}/v_{\rm vir}\sim 1.5$   for $v_{\rm vir}\sim 100{\rm\,km\,s}^{-1}$. 
This result is consistent with the one from weak lensing by \citet{reyes_etal12} when we correct for the difference in our definition of $r_{\rm vir}$.  \citet{dutton_etal10} 
find a curve $v_{\rm rot}/v_{\rm vir}$ vs.
$v_{\rm vir}$ with the same shape and the maximum in the same position but
their values of $v_{\rm rot}/v_{\rm vir}$  are systematically lower (their maximum value for $v_{\rm rot}/v_{\rm vir}$ is closer to 1.1).

In Fig.~3, we have also compared our results to those obtained with the abundance-matching method by \citet{papastergis_etal11}  (solid blue curve; the same approach
has also been explored by \citealp{trujillo_etal11}).
\citet{papastergis_etal11} have considered the galaxy velocity function computed by using the HI rotation speed for disc galaxies and $\sigma\sqrt{2}$ for elliptical galaxies where $\sigma$ is the stellar velocity dispersion at $1/8$ effective radii and they have matched this velocity function to the virial velocity function of DM haloes.
Given that the two methods are totally unrelated, the broad agreement of the red curve and the blue curve is quite significant. Furthermore, a small deviation is not unexpected, since rotation speeds measured from the width of the HI line are generally slightly larger than $v_{\rm opt}$ \citep{dutton_etal10}.

Finally, we recall that, in \citet{tonini_etal06}, we had computed the spin parameter of spiral galaxies by using PSS's universal RC (automatically in agreement with the TF relation) in conjunction
with the disc - halo mass relation from abundance matching (also see \citealp{shankar_etal06}). Our article shows that the approach of these studied was well grounded.

\section{LF and TF relation: reconciled at last}

In Section~2, we have used our analysis of disc RCs to determine the relation between $v_{\rm rot}$ and $v_{\rm vir}$ (the red curve in Fig.~3).
Now, we use this relation to transform the  $M_I$ - $v_{\rm vir}$ relation from abundance matching (the dashed curve in Fig.~1) into a relation
between $M_I$ and $v_{\rm opt}$.
 This relation is shown by the red curve in Fig.~4 and is in excellent agreement with the TF data points (black symbols; they are the same in Fig.~1 and Fig.~4).
 Our conclusion is that  we do not encounter any problem at reproducing the LF and the TF relation simultaneously when the dependence of  $v_{\rm opt}$
 on $v_{\rm vir}$ is properly accounted for.
 
 \begin{figure}
\begin{center}
\includegraphics[scale=0.75]{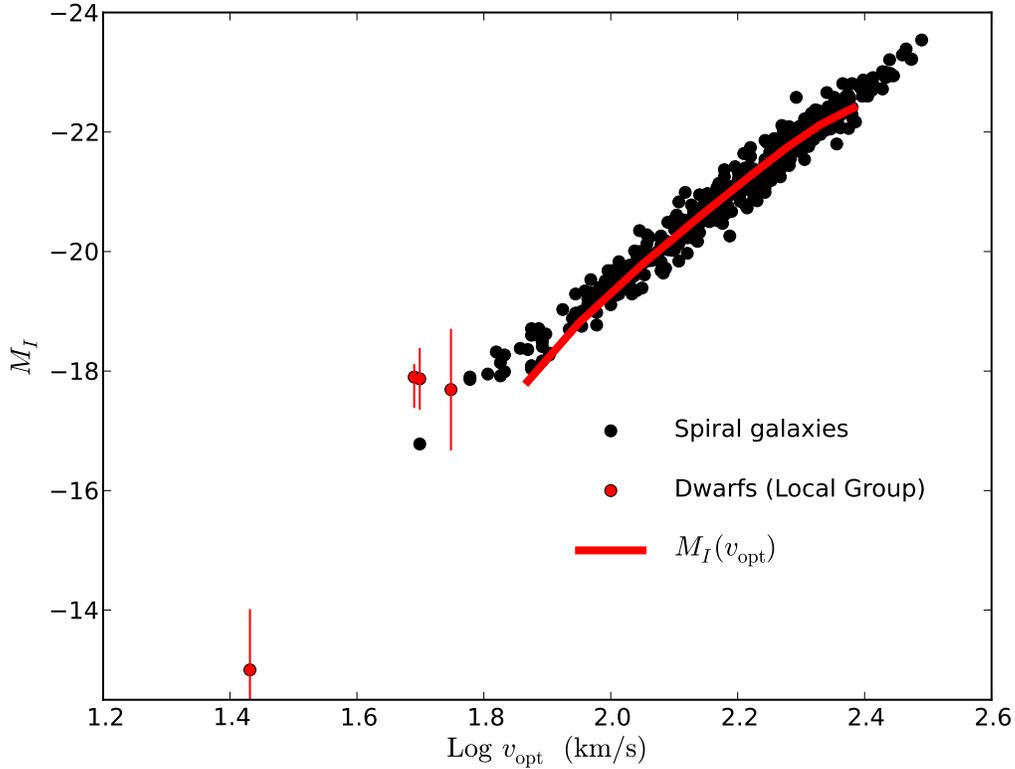} 
\end{center}
\label{stripping_vs_radius}
\caption{
Black symbols: TF data points (same as in Fig.~1).
Red symbols: four dwarf galaxies inserted to extend the TF relation to $M_I\sim -13$ (same as in Fig.~1).
Red curve: the result of converting the $M_I$-$v_{\rm vir}$ relation (the dashed line in Fig.~1) 
 into an $M_I$-$v_{\rm opt}$ relation by using the $v_{\rm rot}$-$v_{\rm vir}$ relation that we extract from our
analysis of observed rotation curves (the red curve in Fig.~3).}
\end{figure}

Our result is established in the magnitude range $-22<M_I<-18$. LF data (from which we derive the dashed curve in Fig.~1)
exist down to $M_I\sim-14$ but
we still lack a statistically significant dwarf-galaxy sample with high-quality RCs that may allow us to extend our analysis at $M_I>-18$, 
although progress in this direction is being made (e.g. the 30-galaxy sample by \citealp{martinsson_etal13}). Even with these uncertainties,
there is evidence that  the TF relation may bend at low luminosities (e.g.  \citealp{trujillo_etal11} and our lowest-luminosity dwarf-galaxy data point).

The quality of the agreement at $-22<M_I<-18$ is also linked to the consistency of our procedure. The $v_{\rm opt}/v_{\rm vir}$ that we use to pass from $v_{\rm vir}$ to 
$v_{\rm opt}$ is measured from the same RCs from which we extract the TF data points.
Furthermore, our choice to work with $I$-band luminosity throughout the article, both for the LF and TF relation, avoids introducing the uncertainty of stellar mass-to-light ratios

At high luminosities, the red curve shows a hint of flattening, which is not seen in TF data points, but this happens because the abundance-matching relation (the black dashes)
is computed from the total LF, which is dominated by early-type galaxies at high luminosity, while TF relation is shown for spiral galaxies only.
The implication is that, for a same $v_{\rm vir}$, spiral galaxies have a higher $I$-band luminosity than elliptical ones.
This result is in agreement with \citet{trujillo_etal11}, and also
with \citet{wojtak_mamon13}, who find that, for a same $M_{\rm halo}$, late-type galaxies have higher $M_\star/M_{\rm halo}$,
where $M_\star$ is the stellar mass.
\citealp{dutton_etal10} reach the same conclusion when making the comparison at a fixed $M_\star$

The dependence of $v_{\rm opt}/v_{\rm vir}$ on $v_{\rm vir}$  is of paramount importance to explain how it is possible to reconcile TF relation (a single power-law)
with the LF, which has a break at the characteristic luminosity $L_\star$. 
This dependence arises from the strong trend in $M_\star/M_{\rm halo}$ with halo mass.
This can be seen both through the disc/halo decomposition of the RCs (PSS) and through the results of abundance matching, stellar-kinematics, and weak-lensing studies
(see \citealp{papastergis_etal12},  where we also compare the results of many different authors).

\bibliographystyle{mn2e}

\bibliography{ref_av}

\begin{thebibliography}{}

\bibitem[\protect\citeauthoryear{{Dwarf spheroidal galaxy kinematics and spiral
  galaxy scaling laws}}{por}{}]{portinari_salucci10}


\bibitem[\protect\citeauthoryear{{Brook}, {Stinson}, {Gibson}, {Wadsley} \&
  {Quinn}}{{Brook} et~al.}{2012}]{brook_etal12}
{Brook} C.~B.,  {Stinson} G.,  {Gibson} B.~K.,  {Wadsley} J.,    {Quinn} T.,
  2012, \mnras, 424, 1275

\bibitem[\protect\citeauthoryear{{Bryan} \& {Norman}}{{Bryan} \&
  {Norman}}{1998}]{bryan_norman98}
{Bryan} G.~L.,  {Norman} M.~L.,  1998, \apj, 495, 80

\bibitem[\protect\citeauthoryear{{Burkert}}{{Burkert}}{1995}]{burkert95}
{Burkert} A.,  1995, \apjl, 447, L25

\bibitem[\protect\citeauthoryear{{Conroy}, {Wechsler} \& {Kravtsov}}{{Conroy}
  et~al.}{2007}]{conroy_etal07}
{Conroy} C.,  {Wechsler} R.~H.,    {Kravtsov} A.~V.,  2007, \apj, 668, 826

\bibitem[\protect\citeauthoryear{{Dutton}, {Conroy}, {van den Bosch}, {Prada}
  \& {More}}{{Dutton} et~al.}{2010}]{dutton_etal10}
{Dutton} A.~A.,  {Conroy} C.,  {van den Bosch} F.~C.,  {Prada} F.,    {More}
  S.,  2010, \mnras, 407, 2

\bibitem[\protect\citeauthoryear{{Flores} \& {Primack}}{{Flores} \&
  {Primack}}{1994}]{flores_primack94}
{Flores} R.~A.,  {Primack} J.~R.,  1994, \apjl, 427, L1

\bibitem[\protect\citeauthoryear{{Fukugita}, {Shimasaku} \&
  {Ichikawa}}{{Fukugita} et~al.}{1995}]{fukugita_etal95}
{Fukugita} M.,  {Shimasaku} K.,    {Ichikawa} T.,  1995, "Publ. of the Astron.
  Soc. of the Pacific", 107, 945

\bibitem[\protect\citeauthoryear{{Geen}, {Slyz} \& {Devriendt}}{{Geen}
  et~al.}{2013}]{geen_etal13}
{Geen} S.,  {Slyz} A.,    {Devriendt} J.,  2013, \mnras, 429, 633

\bibitem[\protect\citeauthoryear{{Governato}, {Brook}, {Mayer}, {Brooks},
  {Rhee}, {Wadsley}, {Jonsson}, {Willman}, {Stinson}, {Quinn} \&
  {Madau}}{{Governato} et~al.}{2010}]{governato_etal10}
{Governato} F.,  {Brook} C.,  {Mayer} L.,  {Brooks} A.,  {Rhee} G.,  {Wadsley}
  J.,  {Jonsson} P.,  {Willman} B.,  {Stinson} G.,  {Quinn} T.,    {Madau} P.,
  2010, \nat, 463, 203

\bibitem[\protect\citeauthoryear{{Guo}, {White}, {Li} \&
  {Boylan-Kolchin}}{{Guo} et~al.}{2010}]{guo_etal10}
{Guo} Q.,  {White} S.,  {Li} C.,    {Boylan-Kolchin} M.,  2010, \mnras, 404,
  1111

\bibitem[\protect\citeauthoryear{{Heyl}, {Hernquist} \& {Spergel}}{{Heyl}
  et~al.}{1995}]{heyl_etal95}
{Heyl} J.~S.,  {Hernquist} L.,    {Spergel} D.~N.,  1995, \apj, 448, 64

\bibitem[\protect\citeauthoryear{{Kauffmann}, {White} \&
  {Guiderdoni}}{{Kauffmann} et~al.}{1993}]{kauffmann_etal93}
{Kauffmann} G.,  {White} S.~D.~M.,    {Guiderdoni} B.,  1993, \mnras, 264, 201

\bibitem[\protect\citeauthoryear{{Klypin}, {Trujillo-Gomez} \&
  {Primack}}{{Klypin} et~al.}{2011}]{klypin_etal11}
{Klypin} A.~A.,  {Trujillo-Gomez} S.,    {Primack} J.,  2011, \apj, 740, 102

\bibitem[\protect\citeauthoryear{{Leauthaud}, {Finoguenov}, {Kneib}, {Taylor},
  {Massey}, {Rhodes}, {Ilbert}, {Bundy}, {Tinker}, {George} \&
  {Capak}}{{Leauthaud} et~al.}{2010}]{leauthaud_etal10}
{Leauthaud} A.,  {Finoguenov} A.,  {Kneib} J.-P.,  {Taylor} J.~E.,  {Massey}
  R.,  {Rhodes} J.,  {Ilbert} O.,  {Bundy} K.,  {Tinker} J.,  {George} M.~R.,
   {Capak} P.,  2010, \apj, 709, 97

\bibitem[\protect\citeauthoryear{{Mandelbaum}, {Seljak}, {Kauffmann}, {Hirata}
  \& {Brinkmann}}{{Mandelbaum} et~al.}{2006}]{mandelbaum_etal06}
{Mandelbaum} R.,  {Seljak} U.,  {Kauffmann} G.,  {Hirata} C.~M.,    {Brinkmann}
  J.,  2006, \mnras, 368, 715

\bibitem[\protect\citeauthoryear{{Martinsson}, {Verheijen}, {Westfall},
  {Bershady}, {Andersen} \& {Swaters}}{{Martinsson}
  et~al.}{2013}]{martinsson_etal13}
{Martinsson} T.~P.~K.,  {Verheijen} M.~A.~W.,  {Westfall} K.~B.,  {Bershady}
  M.~A.,  {Andersen} D.~R.,    {Swaters} R.~A.,  2013, \aap, 557, A131

\bibitem[\protect\citeauthoryear{{Mathewson}, {Ford} \& {Buchhorn}}{{Mathewson}
  et~al.}{1992}]{mathewson_etal92}
{Mathewson} D.~S.,  {Ford} V.~L.,    {Buchhorn} M.,  1992, \apjs, 81, 413

\bibitem[\protect\citeauthoryear{{Montero-Dorta} \& {Prada}}{{Montero-Dorta} \&
  {Prada}}{2009}]{monterodorta_prada09}
{Montero-Dorta} A.~D.,  {Prada} F.,  2009, \mnras, 399, 1106

\bibitem[\protect\citeauthoryear{{Moore}}{{Moore}}{1994}]{moore94}
{Moore} B.,  1994, \nat, 370, 629

\bibitem[\protect\citeauthoryear{{More}, {van den Bosch}, {Cacciato}, {Skibba},
  {Mo} \& {Yang}}{{More} et~al.}{2011}]{more_etal11}
{More} S.,  {van den Bosch} F.~C.,  {Cacciato} M.,  {Skibba} R.,  {Mo} H.~J.,
   {Yang} X.,  2011, \mnras, 410, 210

\bibitem[\protect\citeauthoryear{{Navarro}, {Frenk} \& {White}}{{Navarro}
  et~al.}{1997}]{navarro_etal97}
{Navarro} J.~F.,  {Frenk} C.~S.,    {White} S.~D.~M.,  1997, \apj, 490, 493

\bibitem[\protect\citeauthoryear{{Papastergis}, {Cattaneo}, {Huang},
  {Giovanelli} \& {Haynes}}{{Papastergis} et~al.}{2012}]{papastergis_etal12}
{Papastergis} E.,  {Cattaneo} A.,  {Huang} S.,  {Giovanelli} R.,    {Haynes}
  M.~P.,  2012, \apj, 759, 138

\bibitem[\protect\citeauthoryear{{Papastergis}, {Martin}, {Giovanelli} \&
  {Haynes}}{{Papastergis} et~al.}{2011}]{papastergis_etal11}
{Papastergis} E.,  {Martin} A.~M.,  {Giovanelli} R.,    {Haynes} M.~P.,  2011,
  \apj, 739, 38

\bibitem[\protect\citeauthoryear{{Persic} \& {Salucci}}{{Persic} \&
  {Salucci}}{1995}]{persic_salucci95}
{Persic} M.,  {Salucci} P.,  1995, \apjs, 99, 501

\bibitem[\protect\citeauthoryear{{Persic}, {Salucci} \& {Stel}}{{Persic}
  et~al.}{1996}]{persic_etal96}
{Persic} M.,  {Salucci} P.,    {Stel} F.,  1996, \mnras, 281, 27

\bibitem[\protect\citeauthoryear{{Pontzen} \& {Governato}}{{Pontzen} \&
  {Governato}}{2012}]{pontzen_governato12}
{Pontzen} A.,  {Governato} F.,  2012, \mnras, 421, 3464

\bibitem[\protect\citeauthoryear{{Reyes}, {Mandelbaum}, {Gunn}, {Nakajima},
  {Seljak} \& {Hirata}}{{Reyes} et~al.}{2012}]{reyes_etal12}
{Reyes} R.,  {Mandelbaum} R.,  {Gunn} J.~E.,  {Nakajima} R.,  {Seljak} U.,
  {Hirata} C.~M.,  2012, \mnras, 425, 2610

\bibitem[\protect\citeauthoryear{{Salucci}, {Lapi}, {Tonini}, {Gentile},
  {Yegorova} \& {Klein}}{{Salucci} et~al.}{2007}]{salucci_etal07}
{Salucci} P.,  {Lapi} A.,  {Tonini} C.,  {Gentile} G.,  {Yegorova} I.,
  {Klein} U.,  2007, \mnras, 378, 41

\bibitem[\protect\citeauthoryear{{Salucci}, {Wilkinson}, {Walker}, {Gilmore},
  {Grebel}, {Koch}, {Frigerio Martins} \& {Wyse}}{{Salucci}
  et~al.}{2012}]{salucci_etal12}
{Salucci} P.,  {Wilkinson} M.~I.,  {Walker} M.~G.,  {Gilmore} G.~F.,  {Grebel}
  E.~K.,  {Koch} A.,  {Frigerio Martins} C.,    {Wyse} R.~F.~G.,  2012, \mnras,
  420, 2034

\bibitem[\protect\citeauthoryear{{Shankar}, {Lapi}, {Salucci}, {De Zotti} \&
  {Danese}}{{Shankar} et~al.}{2006}]{shankar_etal06}
{Shankar} F.,  {Lapi} A.,  {Salucci} P.,  {De Zotti} G.,    {Danese} L.,  2006,
  \apj, 643, 14

\bibitem[\protect\citeauthoryear{{Swaters}, {Verheijen}, {Bershady} \&
  {Andersen}}{{Swaters} et~al.}{2003}]{swaters_etal03}
{Swaters} R.~A.,  {Verheijen} M.~A.~W.,  {Bershady} M.~A.,    {Andersen} D.~R.,
   2003, \apjl, 587, L19

\bibitem[\protect\citeauthoryear{{Teyssier}, {Pontzen}, {Dubois} \&
  {Read}}{{Teyssier} et~al.}{2013}]{teyssier_etal13}
{Teyssier} R.,  {Pontzen} A.,  {Dubois} Y.,    {Read} J.~I.,  2013, \mnras,
  429, 3068

\bibitem[\protect\citeauthoryear{{Tonini}, {Lapi}, {Shankar} \&
  {Salucci}}{{Tonini} et~al.}{2006}]{tonini_etal06}
{Tonini} C.,  {Lapi} A.,  {Shankar} F.,    {Salucci} P.,  2006, \apjl, 638, L13

\bibitem[\protect\citeauthoryear{{Trujillo-Gomez}, {Klypin}, {Primack} \&
  {Romanowsky}}{{Trujillo-Gomez} et~al.}{2011}]{trujillo_etal11}
{Trujillo-Gomez} S.,  {Klypin} A.,  {Primack} J.,    {Romanowsky} A.~J.,  2011,
  \apj, 742, 16

\bibitem[\protect\citeauthoryear{{Wojtak} \& {Mamon}}{{Wojtak} \&
  {Mamon}}{2013}]{wojtak_mamon13}
{Wojtak} R.,  {Mamon} G.~A.,  2013, \mnras, 428, 2407

\bibitem[\protect\citeauthoryear{{Yegorova}, {Pizzella} \&
  {Salucci}}{{Yegorova} et~al.}{2011}]{yegorova_etal11}
{Yegorova} I.~A.,  {Pizzella} A.,    {Salucci} P.,  2011, \aap, 532, A105

\bibitem[\protect\citeauthoryear{{Yegorova} \& {Salucci}}{{Yegorova} \&
  {Salucci}}{2007}]{yegorova_salucci07}
{Yegorova} I.~A.,  {Salucci} P.,  2007, \mnras, 377, 507

\end{thebibliography}

\label{lastpage}
\end{document}